\documentclass[pra,notitlepage,nofootinbib,twocolumn]{revtex4-1}
\usepackage{amsmath}
\usepackage{amssymb}
\usepackage{graphicx}
\usepackage[tight]{subfigure}
\usepackage{enumitem}
\usepackage{soul}
\usepackage{pifont}
\usepackage{lipsum}
\usepackage{xcolor}

\usepackage[english]{babel}
\makeatletter
\def\bbl@set@language#1{%
  \edef\languagename{%
    \ifnum\escapechar=\expandafter`\string#1\@empty
    \else\string#1\@empty\fi}%
  \@ifundefined{babel@language@alias@\languagename}{}{%
    \edef\languagename{\@nameuse{babel@language@alias@\languagename}}%
  }%
  \select@language{\languagename}%
  \expandafter\ifx\csname date\languagename\endcsname\relax\else
    \if@filesw
      \protected@write\@auxout{}{\string\select@language{\languagename}}%
      \bbl@for\bbl@tempa\BabelContentsFiles{%
        \addtocontents{\bbl@tempa}{\xstring\select@language{\languagename}}}%
      \bbl@usehooks{write}{}%
    \fi
  \fi}
\newcommand{\DeclareLanguageAlias}[2]{%
  \global\@namedef{babel@language@alias@#1}{#2}%
}
\makeatother
\DeclareLanguageAlias{en}{english}

\definecolor{gr}{rgb}{0,0.82,0.18}

\newcommand{\up}{\mathord{\uparrow}}
\newcommand{\dn}{\mathord{\downarrow}}
\newcommand{\I}{\mathbb{I}}
\newcommand{\tr}{\text{tr}}

\usepackage{tikz}
\usetikzlibrary{calc,fadings,decorations.pathreplacing,shapes,shapes.multipart,arrows,shapes.misc,intersections,positioning}

\newcommand{\zst}{
  \begin{tikzpicture}[baseline={([yshift=-.6ex]current  bounding  box.center)}, scale = 0.5, every node/.style={scale = 1}]
    \draw (0,0) circle (0.15);
    \node () at (0,0) {};
  \end{tikzpicture}
}
\newcommand{\ost}{
  \begin{tikzpicture}[baseline={([yshift=-.6ex]current  bounding  box.center)}, scale = 0.5, every node/.style={scale = 1}]
    \fill (0,0) circle (0.15);
    \node () at (0,0) {};
  \end{tikzpicture}
}

\usepackage{bm}

\usepackage[colorlinks=true,citecolor=blue,linkcolor=magenta]{hyperref}

\graphicspath{{./}{./images/}}

\begin{document}

\title{Diffusive scaling of R\'enyi entanglement entropy}

\author{Tianci Zhou}
\email{tzhou@kitp.ucsb.edu}
\affiliation{Kavli Institute for Theoretical Physics, University of California, Santa Barbara, CA 93106, USA}
\author{Andreas W.W. Ludwig}
\affiliation{Department of Physics, University of California, Santa Barbara, CA 93106, USA}
 
\date{\today}

\begin{abstract}
Recent studies found that the diffusive transport of conserved quantities in non-integrable many-body systems has an imprint on quantum entanglement: while the von Neumann entropy of a state grows linearly in time $t$ under a global quench, all $n$th R\'enyi entropies with $n > 1$ grow with a diffusive scaling $\sqrt{t}$. 
To understand this phenomenon, we introduce an amplitude $A(t)$, which is the overlap of the time-evolution operator $U(t)$ of the entire system with the tensor product of the two evolution operators of the subsystems of a spatial bipartition. As long as $|A(t)| \ge e^{-\sqrt{Dt}}$, which we argue holds true for generic diffusive non-integrable systems, all $n$th R\'enyi entropies with $n >1$ (annealed-averaged over initial product states) are bounded from above
by $\sqrt{t}$. We prove the following inequality for the disorder average of the amplitude, $\overline{|A(t)|} \ge e^{ - \sqrt{Dt}} $, in a local spin-$\frac{1}{2}$ random circuit with a $\text{U}(1)$ conservation law by mapping to the survival probability of a symmetric exclusion process. Furthermore, we numerically show that the typical decay behaves asymptotically, for long times, as $|A(t)| \sim e^{ - \sqrt{Dt}} $ in the same random circuit as well as in a prototypical non-integrable model with diffusive energy transport
but no disorder.
\end{abstract}

\maketitle

\section{Introduction}
\label{Section-Introduction}


Many-body systems that thermalize often possess a globally conserved quantity. 
Typical examples are non-integrable systems with a time-independent Hamiltonian where the conserved quantity is energy. 
The conserved quantity could also be spin or charge in systems with time-dependent Hamiltonians, such as Floquet systems (subject to a time-periodic drive), or systems subject to time-dependent noise. These conserved charges often relax diffusively.
Diffusive systems are those in which time-dependent correlation functions such as $\langle  Z_x(t) Z_0 \rangle $, for a conserved charge density $Z_x$ at position $x$, obey a diffusion equation at coarse grained scales. This therefore generates a hydrodynamic tail in such a correlation function that scales as $\frac{1}{\sqrt{t}}$\cite{kadanoff_hydrodynamic_1963,kirkpatrick_long-time_2002}. There has been also a lot of effort trying to understand the fluctuations on top of the classical diffusion from the point of view of effective field theory (for a recent review see e.g. Ref.~\cite{glorioso_lectures_2018}).

In this work, we focus on the imprint of diffusive transport on dynamical quantities that cannot
be accessed by local correlation functions. For instance the study of the out-of-time-ordered correlator (OTOC) in the context of quantum chaos reveals how the support of an initially local Heisenberg operator $O(t)$ grows with time. In a generic system without any conservation law, the support roughly speaking grows ballistically with the ``butterfly velocity'' $v_B$\cite{nahum_operator_2018,von_keyserlingk_operator_2018}. In contrast, in the presence of a conserved quantity, the Heisenberg operator $O(t)$ was found to contain a diffusive core of the conserved quantity on top of the ballistic components\cite{khemani_operator_2018,rakovszky_diffusive_2017}, making the spreading slower. See also Ref.~\cite{gullans_entanglement_2019} for open systems.


\begin{figure}[ht]
\centering
\subfigure{
  \label{fig:H_L_H_R}
  \includegraphics[width=0.8\columnwidth]{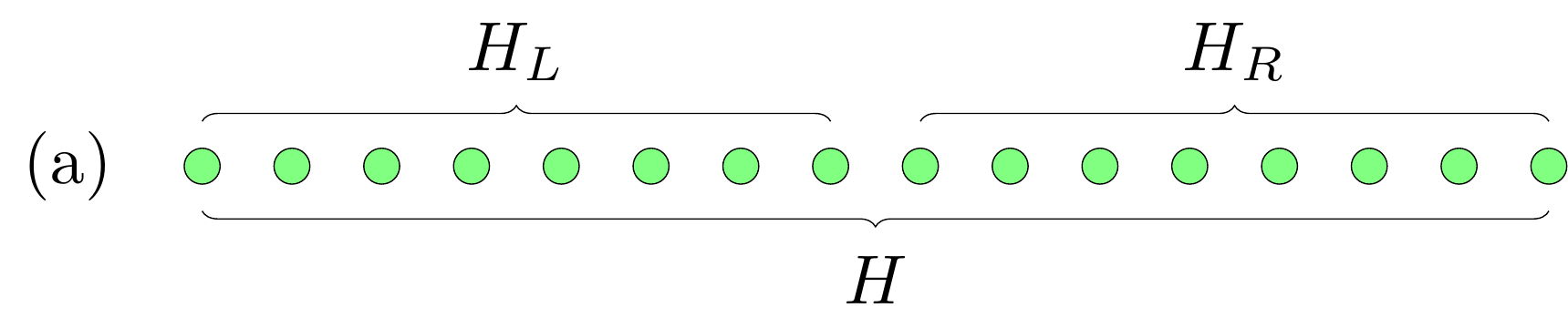}
}\vspace{-10pt}
\subfigure{
  \label{fig:kim_huse}	
  \includegraphics[width=\columnwidth]{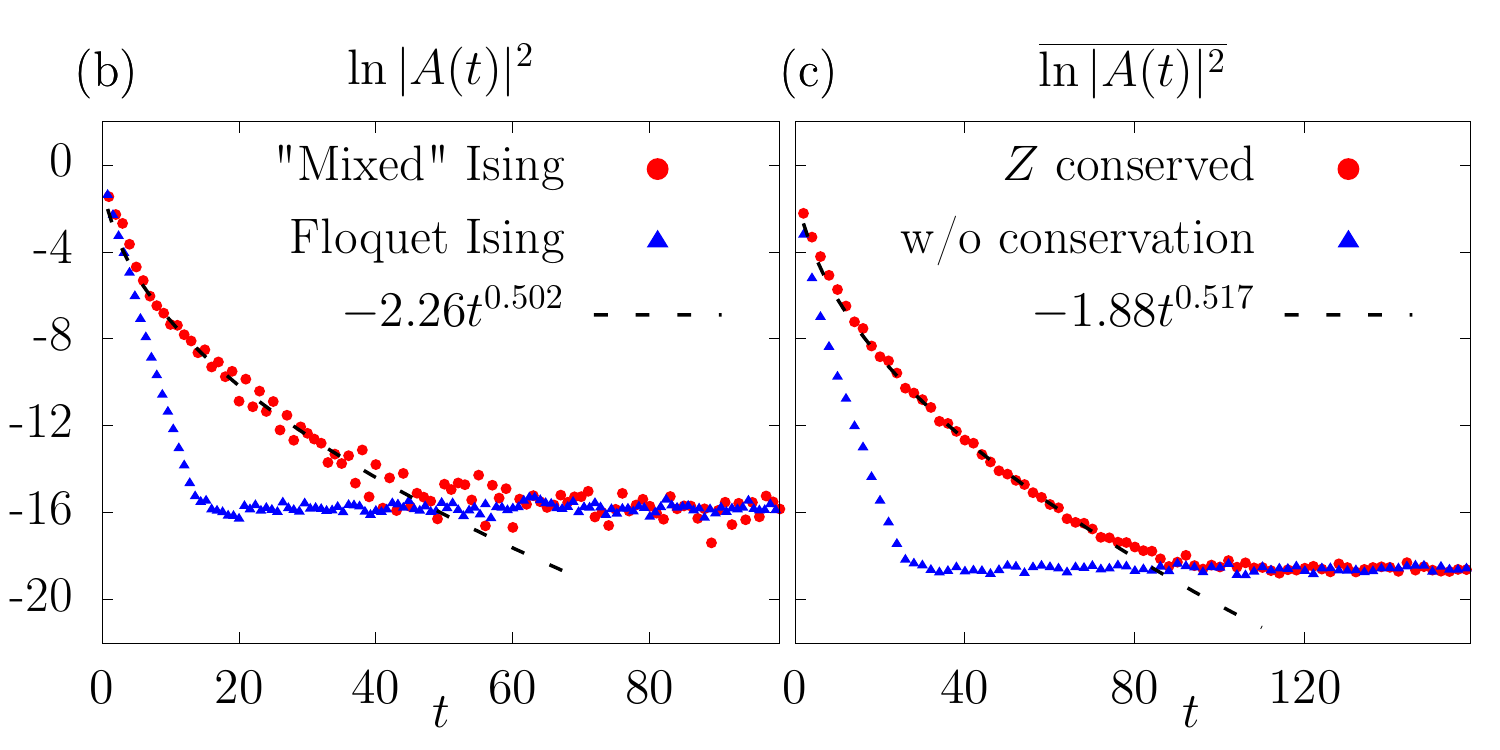}
}
\subfigure{
  \label{fig:rucZ}	
}
\caption{
Numerical evidence for the scaling of $A(t)$  proposed in Eq. (\ref{eq:A_proposal}). (a) The spatial bipartition of the total system. $H$ is the Hamiltonian of the whole system, $H_{L}$ and $H_{R}$ are the Hamiltonians for the left and right semi-infinite parts. (b) $\ln |A(t)|^2$ decreases, respectively, as $-\sqrt{t}$ and $-t$ for a "mixed" Ising 
Hamiltonian evolution (i.e., with both transverse and longitudinal fields) and its Floquet version. System size $L = 22$. (c) The random average (100 samples) $\overline{ \ln |A(t)|^2}$ decreases as $-\sqrt{t}$ and $-t$ in random circuits with and without a ${\rm U}(1)$ conserved charge, respectively. System size $L = 26$.}
\label{fig:headline}
\end{figure}


Perhaps more striking is the scaling of quantum entanglement from a quenched state. A few years ago, Ref.~\onlinecite{kim_ballistic_2013} concluded that for a Hamiltonian evolution (i.e. where energy is conserved) the entanglement spreads ballistically despite the energy diffusion. More precisely, this work provided numerical evidence for a linear growth in time of the {\it von Neumann} entropy $S_A(t) = -\tr( \rho_A(t) \ln \rho_A(t) ) $
of the reduced density matrix $\rho_A(t)$ of a subsystem $A$, time-evolved from a quenched initial state. { Before this question was investigated in detail, it was commonly believed that the {\it R\'enyi} entanglement entropies $S_n(t) = \frac{-1}{n- 1} \ln \tr \rho_A^n(t)$ also grow linearly in time. 
Evidence that may have appeared to support this claim came from systems without any conserved quantities, e.g. quantum quenches in random unitary circuits\cite{oliveira_generic_2007, znidaric_exact_2008, nahum_quantum_2017,zhou_emergent_2019,chan_solution_2018,von_keyserlingk_operator_2018} and exactly solvable kicked Ising models at self-dual points\cite{bertini_entanglement_2018} which indeed exhibit this behavior. 
However, fairly recent work\cite{rakovszky_sub-ballistic_2019} argued that rather different behavior occurs for the R\'enyi entanglement entropies {\it with $n > 1$} of a {\it Hamiltonian}  
evolution
like Ref.~\onlinecite{kim_ballistic_2013}, i.e. of a time-evolution possessing a conservation law (here the energy).} Despite an initial linear growth, this work found that $S_{n>1}(t)$ grows as $\sqrt{t}$ at long times.

This scaling was attributed to the ``slow decay'' of the largest Schmidt eigenvalue of the time evolved reduced density matrix $\rho_A(t)$, which was numerically shown to decay at long times as $e^{-\sqrt{Dt}}$\cite{rakovszky_sub-ballistic_2019}. This largest eigenvalue will dominate in $S_{n>1}$ as all other Schmidt eigenvalues will decay exponentially in time, i.e. much faster. On the other hand, in the {\it von Neumann} entropy the {\it logarithm} of the reduced density matrix is averaged together with the other faster decaying eigenvalues, thus leading to a loss of the diffusive scaling form. This fact was also utilized in Ref. \onlinecite{huang_dynamics_2019} to obtain a slightly weaker bound, where it was shown that $S_{n>1}(t) \le \mathcal{O}(\sqrt{t \ln t})$ with probability $1 - \epsilon(t)$, where $\epsilon(t)$ decays at least as a power law in $t$.

In the present work, we view the problem from the operator perspective. In particular, cutting the 1d diffusive system into disconnected left and right parts, we introduce as a key object the amplitude $A(t)$ defined as
\begin{equation}
\label{eq:DefOfA}
A(t) = \tr \Bigl ( U(t) U^{\dagger}_{L}(t) U^{\dagger}_{R}(t) \Bigr ) / \tr( \I ) 
\end{equation}
and propose that it behaves as\footnote{Strictly speaking, $A(t)$ should be asymptotically equal to $f(t)  \ e^{-\sqrt{Dt}}$, where $f(t)$ is a function that is bounded from below and above $0 < c_1 < f(t) < c_2 $. }
\begin{equation}
\label{eq:A_proposal}
A(t) \propto e^{-\sqrt{Dt}},
\end{equation}
for sufficiently long times\footnote{I.e. for times $t$ larger than the local thermalization time. In the numerical results of Fig.~\ref{fig:headline}, this time scale is of order unity.} $t$. Here $U(t)$ denotes the evolution operator for the whole 1D system, and $U_L(t)$ and $U_R(t)$ denote the evolution operators of the disconnected left and right subsystems, respectively (the latter two being separated by cutting a single bond connecting them - see Fig. \ref{fig:H_L_H_R}). The division of the trace by the Hilbert space dimension normalizes the amplitude $A(t)$ to be $1$ at $t = 0$, and implies $|A(t)| \leq 1$ at any time $t$.
In a system with time independent Hamiltonian $H$ for the total system, we have $U(t) = e^{ - iH t}$ and $U_{L/R}(t) = \exp( - i H_{L/R} t )$, where $H_L$ and $H_R$ are the Hamiltonians restricted to the left and  right subsystems.

The amplitude $A(t)$ measures to what extent the unitary operator $U(t)$ for the whole system can be approximated by the independent evolutions in the subsystem (left) and its complement (right). Our proposal is that at least in generic non-integrable (chaotic) many-body systems, the decay of amplitude $A(t)$ is controlled by the slowest mode of transport: In such a system with a conservation law, the amplitude decays as $e^{-\sqrt{Dt}}$ as displayed in Eq.~\eqref{eq:A_proposal}, in contrast to the case without conservation laws where we expect the amplitude $A(t)$ to decay exponentially in time. 

The diffusive scaling of the $n$th R\'enyi entropy for $n>1$ is a direct consequence of the asymptotics of $A(t)$. In fact, Eq.~\eqref{eq:A_proposal} implies that the {\it operator} R\'enyi entropy of the time evolution operator $U(t)$, upon left-right bipartioning, is bounded from above by 
$-\ln |A(t)| \sim \sqrt{t}$
(see Sec.~\ref{sec:implies_ee}). A  weaker claim can be made for time evolved states as follows:
We define the `{\it annealed averaged}' entropy of the time-evolved state $|\psi(t)\rangle$ to be $S_{n>1}^{(a)}[|\psi(t)\rangle]=$ $-\frac{1}{n-1} \ln[(e^{- (n-1) S_n})]_{av}$, where $S_n$
on the right hand side denotes the R\'enyi entropy of the state $|\psi(t)\rangle$ 
evolved from a random initial product state. We take the average $[ ... ]_{av}$ over those initial product states {\it inside} the logarithm.
We will show below that Eq.~\eqref{eq:A_proposal} implies that $S_{n>1}^{(a)}$ is also bounded from above by $\sqrt{t}$. 

{ We note that the (second) R\'enyi entanglement entropy 
under time-evolution from a fixed (``quenched'') state
has been experimentally measured by various means\cite{brydges_probing_2019,elben_renyi_2018,daley_measuring_2012,islam_measuring_2015}, albeit the number of coherent degrees of freedom 
is still somewhat limited. Making use of the quantity $A(t)$ that we propose in the present paper [Eq.~\ref{eq:DefOfA} above] can greatly reduce effects of finite system size, and is thus more promising as compared to using the $n>1$ R\'enyi entropies themselves which were considered in the current experiments.}

In the following, we present substantial evidence for the proposal in Eq.~\eqref{eq:A_proposal} and the fact that the bound in this equation is saturated. Namely, we first study $A(t)$ in a random circuit model with a ${\rm U}(1)$ conserved charge, which can be regarded as a minimal model for a non-integrable diffusive system. In contrast to previous works on such random circuits\cite{rakovszky_diffusive_2017,khemani_operator_2018,rakovszky_sub-ballistic_2019}, we are able to dispense of an expansion in terms of a large local Hilbert space dimension, which is known to generate ballistic transport that overshadows the diffusive mode of interest. In particular, we analytically prove that the circuit average of the amplitude $\overline{|A(t)|^2}$ decays slower than $e^{- \sqrt{Dt} }$. We also show numerically that in the random circuit the typical decay of the amplitude, i.e. the decay of  $e^{\overline{\ln|A(t)|^2}}$, scales as $e^{- \sqrt{Dt} }$. Finally, we provide numerical evidence that Eq.~\eqref{eq:A_proposal} holds in a non-random chaotic system with a conserved quantity.


\section{Implications for Quantum Entanglement}
\label{sec:implies_ee}

It turns out that the scaling proposed in Eq.~\eqref{eq:A_proposal} constrains the R\'enyi {\it operator} entanglement entropies of the unitary time evolution operator $U(t)$,  as well as the entanglement entropy of a time-evolved state. 
{\it  Operator} entanglement is a natural concept when we view the operator as a state in the Hilbert space of operators. A familiar example is the thermal field double state $ \frac{1}{Z} \sum_i e^{ - \beta H / 2 } |i \rangle |i \rangle $ constructed from the thermal density matrix $\rho_{\rm th} = \frac{1}{Z} \sum_i  e^{ - \beta H } | i \rangle \langle  i | $. 

In general, a non-zero operator $O = \sum_{ij} O_{ij} | i\rangle  \langle j |$, written in an orthonormal basis $\{ |i \rangle \}$ of the Hilbert space ${\cal H}$ it acts on, is mapped to a normalized state $\frac{1}{\sqrt{\tr( OO^{\dagger} )}} \sum_{ij} O_{ij}  | i \rangle |j\rangle $ in the tensor product Hilbert space ${\cal H}\otimes{\cal H}$. A bipartitioning ${\cal H} = {\cal H}_L\otimes{\cal H}_R$ then leads naturally to a bipartitioning of this tensor product Hilbert space
as ${\cal H}\otimes{\cal H} =$ $\left({\cal H}_L\otimes{\cal H}_L\right)
\otimes$  $\left({\cal H}_R\otimes{\cal H}_R\right)$. In complete analogy with a quantum state in a Hilbert space of states ${\cal H}$, one can define the reduced density matrix of an {\it operator}, when the latter is viewed as an element of the vector space ${\cal H}\otimes {\cal H}$ of operators; this is called the reduced {\it operator} density matrix of the operator $O$. The spectrum of the latter  consists, as in the case of a state, of the set of entanglement eigenvalues $\{\lambda^{\rm op}_i\}$, from which the {\it operator} R\'enyi entropies can be computed in the familiar fashion
\begin{equation}
S^{\rm op}_{n} [O] = -\frac{1}{n - 1} \ln( \sum_i (\lambda^{\rm op}_i)^n  ).
\end{equation}
{ Operator
entanglement has been studied for various purposes ranging from efficiently simulating Heisenberg  operators to probing the onset of quantum chaos\cite{dubail_entanglement_2017,bandyopadhyay_entangling_2005,pizorn_operator_2009,prosen_operator_2007,prosen_third_2008,pizorn_real_2014,wang_barrier_2019,nie_signature_2018,zhou_operator_2017}.}

We now apply this to the time evolution operator, i.e. $O \to U(t)$, so that the eigenvalues become time dependent, $\lambda_i^{\rm op} \to\lambda_i^{\rm op}(t) $.
It turns out that the largest entanglement eigenvalue leads to an upper bound for the absolute value of our amplitude, $|A(t)|$ (See Appendix~\ref{app:op_EE_bd}):
\begin{equation}
\label{eq:A_le_max_lamb}
|A(t) | \le \max_i \left\{ \sqrt{\lambda^{\rm op}_i(t)}\right\}.
\end{equation}
From this we see that if we assume the scaling in Eq.~\eqref{eq:A_proposal}, we obtain
$S^{\rm op}_{\infty} [ U(t) ]  = - \ln \Bigl (\max_i \{ \lambda^{\rm op}_i(t)\}\Bigr )  \le 2 \sqrt{Dt} $. From the following well-known inequalities satisfied by the R\'enyi entropies when $n >1$, $S_{\infty} \le  S_n \le \frac{n }{ n- 1 }  S_{\infty} $, we then obtain an upper bound for the {\it operator} R\'enyi entropies
\begin{equation}
\label{LabelEqOperatorRenyiEntropies}
S^{\rm op}_{n} [ U(t)] \le \frac{2n }{n - 1} \sqrt{Dt},
\qquad
 {\rm when} \ n > 1.
\end{equation}

This is notable compared to the {\it linear} growth of the {\it von Neumann} {\it operator} entropy $S_{1}^{\rm op}[U(t)]$, 
which was numerically observed in Ref.~\onlinecite{zhou_operator_2017} for systems with an (energy) conservation law.
\footnote{The result in Ref.~\onlinecite{zhou_operator_2017} can thus be viewed as an analog for {\it operator} entanglement growth of the observation made in Ref.~\onlinecite{kim_ballistic_2013} for entanglement growth of {\it states}.}

To put this into state language, we consider the R\'enyi entropy of a time-evolved state $|\psi(t) \rangle$ where the initial state is taken to be a product state 
\begin{equation}
  |\psi (0) \rangle = \otimes_{i \in \text{sites}}  |\psi_i \rangle 
\end{equation}
of random on-site states $|\psi_i \rangle  =  \cos \frac{\theta_i}{2} |\up \rangle + \sin \frac{\theta_i }{2} e^{ i \phi_i }  | \dn \rangle$ statistically independent at each site with a uniform measure over solid angles. Since then the random average at each site yields $[|\psi_i \rangle  \langle \psi_i |]_{av} = \frac{1}{2} {\I}_i$, we have
\begin{equation}
\label{eq:AverageInitialStatesAt}
[ \langle  \psi (0) | U_L^{\dagger}(t) U_R^{\dagger}(t)  U(t) | \psi(0) \rangle ]_{av}  = A(t).
\end{equation}

Let the entanglement eigenvalues of the state $|\psi(t) \rangle$ for a half cut (as in Fig. \ref{fig:H_L_H_R}) be $\{\lambda_i^{\psi}(t)\}$. Then 
by a reasoning analogous to that in Eq.~\eqref{eq:A_le_max_lamb} we show at the end of Appendix A that (before averaging)
\begin{equation}
\label{eq:LambdaPsiMaxInequality}
  |\langle  \psi (0) | U_L^{\dagger}(t) U_R^{\dagger}(t)  U(t) | \psi(0) \rangle| \le   \sqrt{\lambda^{\psi}_{\rm max}(t)  }.
\end{equation}

Hence, upon averaging over initial states, we obtain in combination\footnote{For a complex random variable $z$:   
$|[z]_{av}| \leq \sqrt{[|z|^2]_{av}} \leq [\sqrt{|z|^2}]_{av} = [|z|]_{av}$.}
with Eq.~\ref{eq:AverageInitialStatesAt}
\begin{equation}
\label{eq:ALessEqualSqrtPsi}
|A(t)| \le [\sqrt{\lambda^{\psi }_{\rm max} (t)  }]_{av} \le  \sqrt{ [\lambda^{\psi }_{\rm max} (t) ]_{av} }.
\end{equation}

In fact, we can use the amplitude $A(t)$ to obtain a lower bound for the  `{\it annealed averaged}' R\'enyi entropies
of a time-evolved state $|\psi(t)\rangle$, defined above: On general grounds we have for $n>1$:
\begin{align}
[e^{- (n-1) S_n}]_{av} &= 
[\sum_i ( \lambda_i^{\psi}(t) )^n ]_{av}   \notag \\
&\ge
([\sum_i ( \lambda_i^{\psi}(t) )]_{av})^n 
\ge
[\lambda_{\rm max}^{\psi}(t) ]_{av}^n.
\end{align}

The last inequality follows since all $\lambda_i^{\psi} \geq 0$. Then using Eq.~\eqref{eq:ALessEqualSqrtPsi}, Eq.~\eqref{eq:A_proposal} implies the following upper bound for the `{\it annealed averaged}'  $n$th R\'enyi entropies with $n > 1$
\begin{equation}
S^{(a)}_n(t) =\frac{-1}{n-1}
\ln \{[\sum_i ( \lambda_i^{\psi}(t) )^n ]_{av}\} 
\le  \frac{2n }{n - 1} \sqrt{Dt}.
\end{equation}
The singular behavior of the upper bound at $n = 1$ is indicative of the fact that we can not na\"ively take the $n\rightarrow 1$ limit to obtain the von-Neumman entropy at large times. The same applies to the operator entanglement in Eq.~\ref{LabelEqOperatorRenyiEntropies}. This suggests that for diffusive non-integrable systems the large $t$ limit and the replica limit $n \rightarrow 1$ do not commute. 


\section{Random Unitary Circuit with Conservation Law}
\label{sec:rucZ}


\begin{figure}[ht]
\centering
\includegraphics[width=0.9\columnwidth]{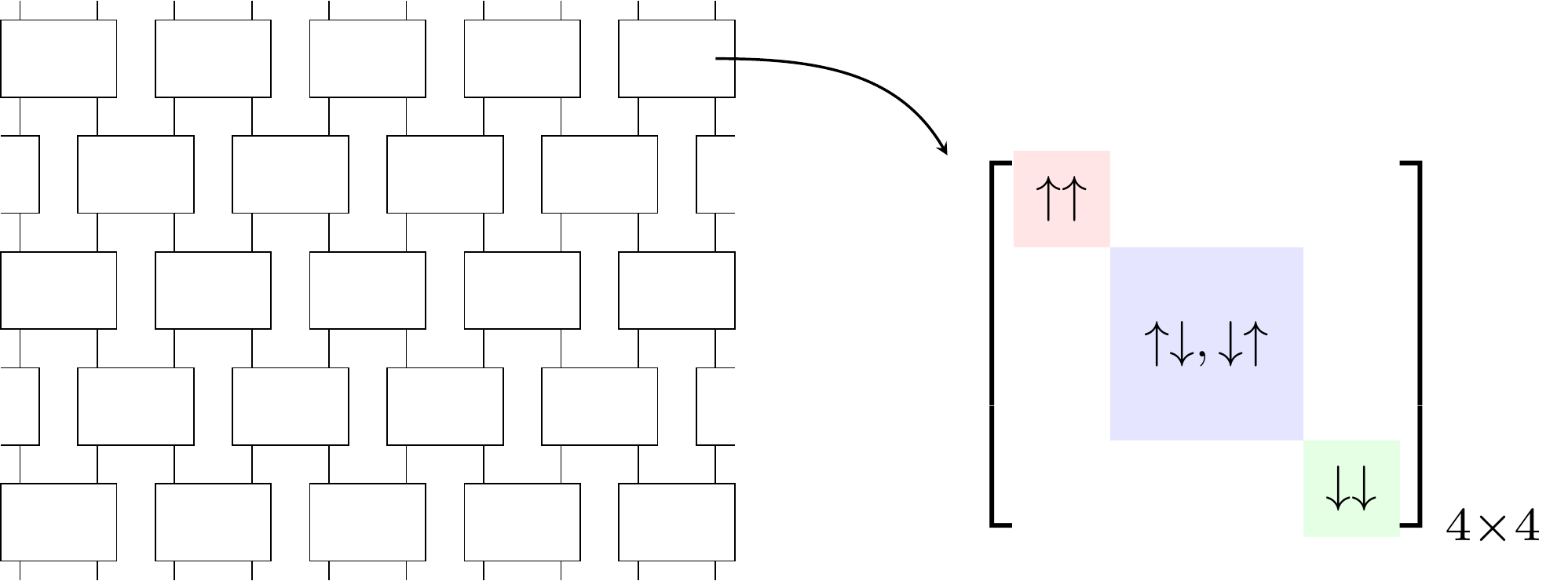}
\caption{Random unitary circuit of a spin-$\frac{1}{2}$ chain with a $\sigma^z$-conservation law. 
The circuit has brick-wall structure so that each gate acts on a nearest neighbor pair of qubits. Each 
gate is a block diagonal matrix that conserves the total $\sigma^z$ spin. Each block is taken to be an independent random matrix from the Haar ensemble.}
\label{fig:rucZ_struc}
\end{figure}

In this section we study analytically a random unitary circuit with a two-dimensional (``single qubit'', or ``spin-1/2'') onsite Hilbert space. The circuit has a brick wall structure to model local interactions, see Fig.~\ref{fig:rucZ_struc}. Such circuits have been used to investigate entanglement\cite{nahum_quantum_2017,skinner_measurement-induced_2018,zhou_emergent_2019} as well as
out-of-time ordered correlation functions\cite{nahum_operator_2018} in chaotic many-body systems with\cite{rakovszky_diffusive_2017,rakovszky_sub-ballistic_2019,khemani_operator_2018} and without\cite{von_keyserlingk_operator_2018} conserved charges.
The variant that we consider here conserves the total
$\sigma^z$ (Pauli-matrix) spin, and each gate represents a $4\times 4$ block diagonal unitary matrix acting on two nearest-neighbor qubits (Fig.~\ref{fig:rucZ_struc}, right panel). We take each block to be randomly sampled from the Haar ensemble. Specifically, the two $1\times 1$ blocks for nearest neighbor qubit configurations $\up \up$ and $\dn\dn $ are two random phases. 

The previous analytic analysis in Refs.~\onlinecite{rakovszky_diffusive_2017,khemani_operator_2018,rakovszky_sub-ballistic_2019} uses a model with a $2 \times q$-dimensional onsite Hilbert space, in which the degrees of freedom of the $q$-dimensional factor (``spectators'') do not carry information about the conserved quantity, and there is a $4q^2 \times 4q^2$ unitary matrix for each nearest neighbor two-site gate. A limit of infinite onsite Hilbert space dimension $q \to \infty$
was then employed in those works to enable an analytic treatment.
However, to see the claimed $e^{-\sqrt{Dt}}$ decay of $A(t)$ in this kind of system one needs to remain in the $q=1$ limit.
This is because the ``internal''("spectator") Hilbert space of dimension $q$ at each site creates ballistic modes when the system scrambles, which give rise to a linear growth term in $S_{n>1}(t)$ even at late time, while the diffusion physics arising from the conservation law gives rise only to a {\it subleading} $\sqrt{t}$ correction - see Ref.~\onlinecite{rakovszky_sub-ballistic_2019}.
This behavior would only imply a bound $A(t) \le e^{-v t }$ for the quantity $A(t)$ at long times in the large-$q$ limit, where $v$ is proportional to the rate of growth of the corresponding R\'enyi entropy.

\begin{figure}[ht]
\centering
\includegraphics[width=\columnwidth]{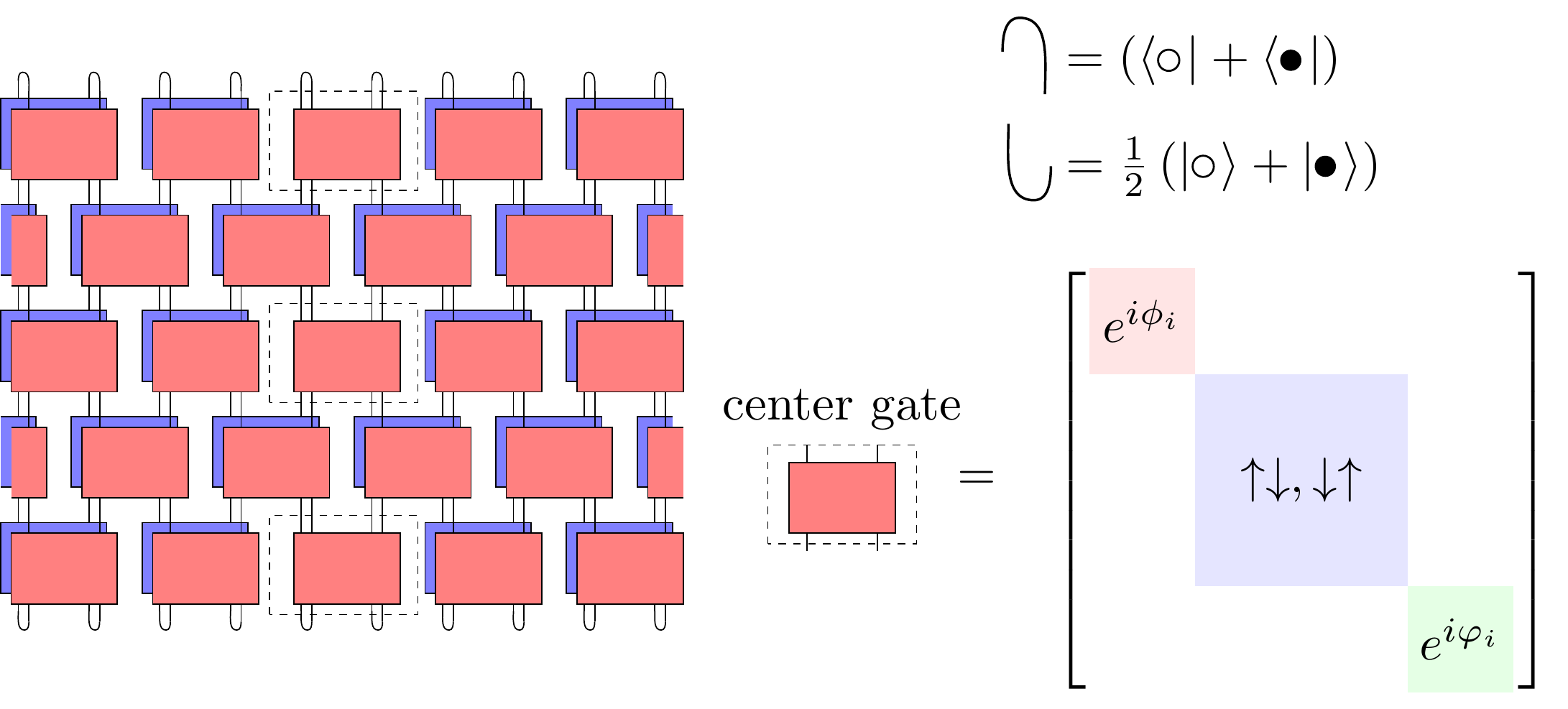}
\caption{The amplitude $A(t)$ shown in terms of circuit diagrams. The red gates (front layer) represent the time evolution operator $U(t)$ for the whole system, the blue gates (back layer) represent the product of the time evolution operators of the left and right half systems, $U_L^{\dagger}(t) U_{R}^{\dagger} (t )$, where the links {\it in the center} between the left and right parts have been removed. The trace operation is understood as bell pair states at both the bottom and the top of the circuit. The unpaired (red) {\it center} gate, which has no (blue) partner in the back layer, is a block diagonal random matrix, where the two $1\times 1$ blocks are two random phases.}
\label{fig:A_rucZ}
\end{figure}

Now we proceed with the random circuit implementation of $A(t)$ for $q=1$ (the case we consider here),
shown in Fig.~\ref{fig:A_rucZ}. Note that Fig.~\ref{fig:A_rucZ} depicts a two-layer circuit: The circuit in
the front layer, denoted by red two-site gates, represents the full evolution $U(t)$. The circuit in the back layer, denoted by blue two-site gates, is identical to the one in the front layer, {\it except} that the {\it center two-site gate is missing}. The circuit in the back layer (blue) thus represents an {\it independent} evolution of the right- and the left-half subsystems, and is described by the evolution operator $U_L^\dagger(t)U_R^\dagger(t)$. The front (red) and back (blue) layers of the random circuit in the Figure thus represent the two ingredients entering into the amplitude $A(t)$ of Eq.~\ref{eq:DefOfA}.

In the following, we will prove for the so-defined circuit (with $q=1$, i.e. with two-dimensional on-site Hilbert space), that the circuit average of the amplitude $A(t)$ satisfies the lower bound
\begin{equation}
\overline{ |A(t)| }\ge e^{-\sqrt{D t} }.
\end{equation}

The amplitude $A(t)$, Eq.~\ref{eq:DefOfA}, starts off at $A(t = 0 ) = 1$, but will in general be a complex number for $t > 0$. In fact, the random evolution above will generate a random phase in  $A(t)$ and so its circuit average vanishes, $\overline{A( t> 0 )} = 0$. The estimation of its magnitude calls for the evaluation of 
the circuit average $\overline{|A|^2}$. This calculation involves a random average over four layers of the circuit rather than over just the two layers shown in Fig.~\ref{fig:A_rucZ}. The random average over four layers produces a many-body problem that contains even more degrees of freedom per site than the original spin-$\frac{1}{2}$ chain. Making things worse, it has a sign problem and lacks a stochastic interpretation
(See Ref.~\cite{daniel_a._rowlands_noisy_2018}, and also the discussion in Ref.~\cite{gullans_entanglement_2019}). It thus remains practically intractable so far. 

We circumvent this problem by isolating the phase and magnitude of $A(t)$.  We focus first on the 2-site unitary {\it center} gates: There are two random phases in a two-site unitary center 
gate of the front layer (red) of the circuit, namely one in each of the two $1\times 1$ blocks. [As mentioned, there is {\it no corresponding} two-site unitary center gate in the back layer (blue) circuit in Fig.~\ref{fig:A_rucZ}.] We denote these phases by $e^{i \phi_i} $ and $e^{i \varphi_i }$ for the $i$th gate among the $\frac{t}{2}$ (red) center two-site gates of the ``front'' circuit (i.e., $i=1, 2, ..., t/2$, where $t$ denotes the discrete time-step). 
We first obtain an estimate of the quantity $\overline{A(t)|_{\phi_i = \varphi_i = 0}}$, in which we have fixed the two phases in each of its $t/2$ ''front'' (red) center two-site gates to unity, while the average is performed over the remaining random parameters of the circuit. 
These parameters include the $2 \times 2$ block in each of the $t/2$ ``front'' (red) center 2-site gates, as well as the matrix elements of all the other front and back 2-site unitary gates.

In the 2-layer unitary structure, $A(t)$ can be compactly written as a Loschmidt echo in the form $A(t) = \langle \psi_f | U_{2\text{-layer}}(t) | \psi_i\rangle $:
Here the operator $ U_{2\text{-layer}}(t)$ represents the 2-layer (front and back) unitary evolution depicted in
Fig.~\ref{fig:A_rucZ}, and is equal to $U(t) \otimes \left ( U^*_L(t)U^*_R(t)\right)$. Due to their special importance after the random 
circuit average (see below), we introduce the following notation for singe-site states of the front and back layers: 
\begin{equation}
\label{eq:def_ef}
| \zst \rangle \equiv |\up \rangle_{\rm front} \otimes |\up  \rangle _{\rm back} \,\, ,\,\,  |\ost \rangle\equiv | \dn \rangle_{\rm front}  \otimes |\dn  \rangle_{\rm back}. 
\end{equation}
We take the initial state $|\psi_i \rangle $ to be a tensor product of such states over sites
\begin{equation}
\label{LabelEqInitialState}
  |\psi_i \rangle = \underset{\rm sites}{\otimes}\frac{1}{2} \left( | \zst \rangle + | \ost \rangle  \right),
\end{equation}
where the factor of $\frac{1}{2}$ will account for the normalization of the amplitude $A(t)$. The final state $|\psi_f\rangle $ is taken to be $ 2^L |\psi_i \rangle$. The initial and final states represent the lines in Fig.~\ref{fig:A_rucZ} that connect at each site the front and back layers at the top and bottom boundaries. 
In the matrix element $A(t) = \langle \psi_f | U_{2\text{-layer}}(t) | \psi_i\rangle $, they therefore implement the trace in the definition, Eq.~\ref{eq:DefOfA}, of the amplitude $A(t)$. 

The random circuit average of $U_{2\text{-layer}}(t)$ has simple effects when acting on states $|\zst\rangle$ and $|\ost\rangle$ (defined in Eq.~\ref{eq:def_ef})
at neighboring sites. Away from the center, where the 2-site `front' (red) and `back' (blue) gates defined above
are ``paired'' with each other, one easily finds 
(see App.~\ref{app:rand_aver_u1})
that upon averaging
\begin{equation}
\label{eq:ssep}
\begin{aligned}
&|\zst \zst \rangle \rightarrow |\zst \zst \rangle,  \quad |\zst\ost\rangle \rightarrow \frac{1}{2} \left( |\zst\ost\rangle + |\ost\zst\rangle  \right )\\ 
&|\ost \ost \rangle \rightarrow |\ost \ost \rangle,  \quad |\ost\zst\rangle \rightarrow \frac{1}{2} \left( |\zst\ost\rangle + |\ost\zst\rangle  \right ).\\
\end{aligned}
\end{equation}
In contrast, the random average of the center 2-site gates, while keeping $|\zst\zst\rangle$ and $|\ost \ost\rangle$ invariant (since we set their random phases to be $1$), completely decimates the remaining two states at the center two sites:
\begin{equation}
\label{eq:decimation}
\begin{aligned}
|\zst\ost\rangle \rightarrow 0 \quad |\ost\zst\rangle \rightarrow 0\\
\end{aligned}
\end{equation}

\begin{figure}[ht]
\centering
\subfigure[]{
  \label{fig:ssep}	
  \includegraphics[width=0.9\columnwidth]{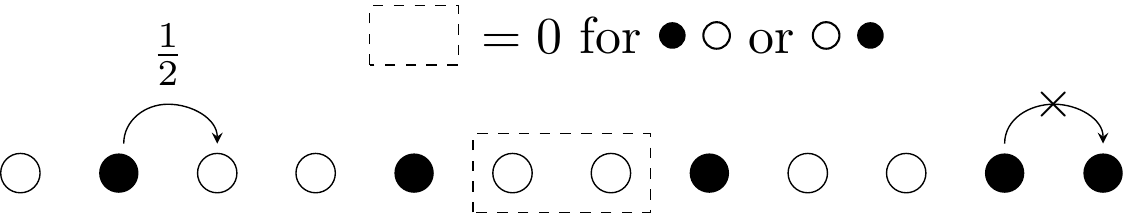}
}
\subfigure[]{
  \label{fig:ssep_l}	
  \includegraphics[width=0.9\columnwidth]{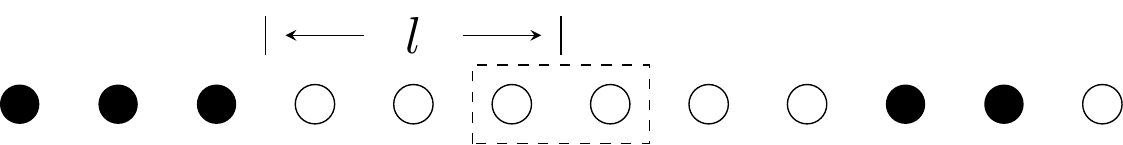}
}
\caption[The random average $\overline{A(t)|_{\phi_i = \varphi_i = 0}}$]{The random average $\overline{A(t)|_{\phi_i = \varphi_i = 0}}$. (a) Spin states are represented in the  $\zst$ and $\ost$ notation. The Figure shows the average effects of applying a random gate. Away from the center, the $\ost$ particle has $\frac{1}{2}$ probability of moving to a neighboring $ \zst $ state. It will stay if the neighboring site is also the $\ost$ state (hard core constraint). Any configuration with $ \ost \zst $ or $ \zst \ost $ at the center will be killed in one step. (b) A domain of $\zst$ with size $l$ symmetric around the center.}
\label{fig:A_ssep}
\end{figure}

We therefore obtain a stochastic process (Fig.~\ref{fig:ssep}) for which
$\overline{A(t)|_{\phi_i = \varphi_i = 0}}$ is the survival probability:
The initial state $|\psi_i \rangle $ is 
an ensemble (equal weight sum) of all the configurations of on-site states $\zst$ and $\ost$ 
appearing with equal probability. On sites away from the center, the transition probability in Eqs.~\eqref{eq:ssep} is that of a symmetric simple exclusion process [SSEP] of $\zst$ and $\ost$ particles\cite{kipnis_scaling_1999,goncalves_simple_2011,derrida_exact_1993,derrida_non-equilibrium_2007}.
They perform a random walk subject to a hard core constraint. One can check that the initial state $|\psi_i \rangle $,
Eq.~\ref{LabelEqInitialState}, is invariant under the `SSEP' rule Eqs.~\eqref{eq:ssep} for each pair of sites {\it not} located at the {\it center}. 
This is a consequence of the unitarity of the gates: Contracting the two-site gates $u$ and $u^*$ between the front and back layers of the circuit as shown
at the bottom of Fig. \ref{fig:A_rucZ}  amounts to the matrix product $u u^{\dagger} = I$, which reproduces the initial state.
The {\it center} 2-site gate breaks this balance of $\zst$ and $\ost$ `particles' implied by the SSEP, by removing the states that contain configurations
$| \zst \ost \rangle $ or $| \ost \zst \rangle $ at the {\it center} every two steps. The overlap of the initial 2-layer state in Eq. \ref{LabelEqInitialState}, time evolved with the 2-layer time evolution operator $U_{2-\text{layer}}(t)$, with $|\psi_f \rangle$ thus counts the  survival probability of the particle configurations in the state $|\psi_i\rangle$. 

Among all, those configurations with a large domain of contiguous $\zst$ or $\ost$ particles spanning the center are more likely to survive in the end. 
In particular, consider an ensemble of configurations containing domains of size $l$ of contiguous states $\zst$ arranged symmetrically around the center, while having equal weight for finding particles $\zst $ and $\ost$ outside (for example, see the configuration in Fig.~\ref{fig:ssep_l}). Due to the diffusion induced by the symmetric exclusion process\cite{kipnis_scaling_1999}, it takes of the order of $l^2$ time steps for a $\ost$ particle to reach the center, before the configuration is decimated. To avoid this fate within $t$ steps, we need $l^2 \gtrsim t$. The only exception is the case where $l = 0$, and we may consider similarly a domain of $\ost$ particles, which works analogously. Combining both cases, the survival probability scales as
\begin{equation}
  \overline{A( t )|_{\phi_i = \varphi_i = 0}} \sim \sum_{l \gtrsim \sqrt{t}}  2^{-l} \sim e^{ - \sqrt{Dt} }.
\end{equation}

Next we relax the constraint of the random phases, and compute the average $\langle \Big|\overline{A( t )|_{\phi_i,\varphi_i}} \Big| \rangle_{\phi_i, \varphi_i} $: The procedure here is to first take the random average given a fixed set of phases $\phi_i, \varphi_i$ (not necessarily $0$), then to take the absolute value, and finally to average over the choices of phases $\phi_i, \varphi_i$. The estimation of $\left|\overline{A( t )|_{\phi_i,\varphi_i}} \right| $ is similar to that of $\overline{A( t )|_{\phi_i = \varphi_i = 0}}$. The only change is the effect of the {\it center} gate, which leads to multiplication with an additional phase at each time step as follows: 
\begin{equation}
\begin{aligned}
&|\zst \zst \rangle \rightarrow e^{i \phi_i } |\zst \zst \rangle  \\
&|\ost \ost \rangle \rightarrow e^{i \varphi_i} |\ost \ost \rangle.  \\
\end{aligned}
\end{equation}
(Recall that $i=1, 2, ..., t/2$ denotes time steps.)

Again, we consider the same ensemble of size-$l$ domains of particles $\zst$  arranged symmetrically around the center, considered above. Before a $\ost$ particle reaches the center, the dynamics is almost the same as discussed above, but now accompanied by a {\it common phase} $e^{ i \sum \phi_i }$ accumulated for each configuration surviving at time $t$. After a $\ost$ particle reaches the center, the configuration will be decimated. Similarly, if we start with a domain of $\ost$ particles, the accumulated common phase is $e^{ i \sum \varphi_i }$. Therefore, the dominant contribution to $|A(t)|$ will be the survival probability of each of the two ensembles times their common phases
\begin{equation}
\begin{aligned}
  \left|\overline{A( t )|_{\phi_i , \varphi_i}} \right|  &= \left|e^{i \sum_i \phi_i } e^{ - \sqrt{D t} }  +  e^{i \sum_i \varphi_i } e^{ - \sqrt{D t} }  \right|  \\
  &\sim e^{ - \sqrt{D t} } \left|\cos \Big(\sum_i (\phi_i - \varphi_i)/ 2  \Big) \right|.
\end{aligned}
\end{equation}
Thus, the average over the random phases, performed {\it after} taking the absolute value, will not affect the $t$ scaling, yielding 
\begin{equation}
\langle \left|\overline{A( t )|_{\phi_i , \varphi_i}} \right| \rangle_{\phi_i, \varphi_i} \sim e^{ - \sqrt{Dt} }. 
\end{equation}
Here the overbar $\overline{( ... )}$ inside denotes the random circuit average (with fixed $\phi_i, \varphi_i$), and the bracket outside denotes the average over the random phases $\phi_i, \varphi_i$. 

Finally, in view of the 
inequalities
$\langle |\overline{A( t )|_{\phi_i, \varphi_i}} | \rangle_{\phi_i, \varphi_i}  \le \overline{|A|} \le \sqrt{ \overline{|A|^2} } $, we have
\begin{equation}
\sqrt{ \overline{|A|^2}}  \ge \overline{|A|} \ge  e^{-\sqrt{Dt} } .
\end{equation}

Let us now discuss the role played by the random circuit average. A single site on two-layers has Hilbert space dimension $4$. The random circuit average allows us to work in the smaller
effective local Hilbert space of dimension $2$, which is spanned by the $|\ost \rangle$ and $|\zst\rangle$ on-site basis defined in Eq.~\ref{eq:def_ef}. This basis produces classical configurations on which the dynamics of $\overline{A(t)}$ has a simple stochastic interpretation. Instead of the disorder average, consider now a single disorder realization of the random circuit: the state $U_{2\text{-layer}} | \psi_i\rangle$ will contain a "classical component" completely expressed in the $|\ost \rangle$ and $|\zst\rangle$ basis and a "noisy component" which allows for onsite states of the form $|\up \rangle_{\rm front} \otimes |\dn \rangle_{\rm back} $ or $|\dn \rangle_{\rm front} \otimes |\up\rangle_{\rm back} $ to appear. Recall that those were excluded from the onsite states $\zst$ and $\ost$ defined in Eq.~\ref{eq:def_ef}. The overlap of the ``noisy component'' with the final state $|\psi_f \rangle $ vanishes, and therefore does not contribute to the amplitude $A(t)$ in the final time-step. But at intermediate time steps the "noisy components" appear to act as a random environment that produces noise in the hopping amplitudes of the $\zst$ and $\ost$ particles. Since the system is still reflection symmetric on average, we expect that in the long time limit, the $\zst$ and $\ost$ particles should have no preferred hopping direction and they will thus generically still diffuse in the hydrodynamic limit. Therefore we expect to recover the $e^{- \sqrt{Dt}}$ scaling for the amplitude $A(t)$ also in a typical realization of the circuit. We also conjecture that such a similar mechanism works in generic (non-integrable) diffusive systems without disorder. 


\section{Numerical Results}


In this section we check the scaling proposed in Eq.~\eqref{eq:A_proposal} numerically for systems with disorder, as well as for chaotic systems without disorder.  

At the end of Sec.~\ref{sec:rucZ}, we argued that diffusive scaling generically holds for a particular realization of the random circuit (with conserved charge). Here we provide numerical evidence for this statement as shown in Fig.~\ref{fig:rucZ}. The figure shows the circuit average $\overline{\ln |A|^2}$ as a function of time for system size $L = 26$. The fitted exponent is very close to $\frac{1}{2}$. This implies that the typical value of $|A|^2$ in the random circuit scales as $e^{- \sqrt{Dt}}$ for $t > t_0$, where $t_0$ is a system size independent constant of order unity. 
For comparison, we computed the same quantity for a random circuit without any conservation law, i.e. taking a full $4\times 4$ random unitary for each gate. The slowest mode then becomes ballistic and we obtain a much faster exponential decay in time (also shown in Fig.~\ref{fig:rucZ}).

Next we examine a clean chaotic system -- the ``mixed field Ising'' Hamiltonian (with both transverse and longitudinal fields) used in Ref.~\onlinecite{kim_ballistic_2013}
\begin{equation}
\label{eq:kim_huse_H}
H = \sum_{i=1}^L h \sigma^z_i + \sum_{i=1}^L g \sigma^x_i + J\sum_{i=1}^{L-1} \sigma^z_{i} \sigma^z_{i+1} 
 - J \sigma^1_z - J \sigma^L_z .
\end{equation}
The conserved quantity is now energy. We choose the parameters $J = 1, g = 0.9045, h = 0.8090$, with which the spectral correlations of the Hamiltonian were confirmed to be those of a corresponding random matrix ensemble\cite{kim_ballistic_2013}. As a result, this system is often regarded as a prototypical example of a diffusive nonintegrable model. The additional boundary fields at site $1$ and $L$ are introduced to reduce finite size effects. We compute $|A(t)|$ where $H_L$ and $H_R$ take the same form as Eq.~\eqref{eq:kim_huse_H} but restricted, respectively, to
the decoupled right and left subsystems of half the length. Fig.~\ref{fig:kim_huse} shows the quantity $\ln |A(t)|^2$ that is seen to decrease as $-\sqrt{t}$, which agrees with our prediction. 

For comparison, we also compute $\ln |A(t)|^2$ for the Floquet version of the model in which energy conservation is absent. The Floquet operator is 
\begin{equation}
U = \exp[ - i \tau ( \sum_{i=1}^L h \sigma^z_i  + J\sum_{i=1}^{L-1} \sigma^z_{i} \sigma^z_{i+1}   )  ] \exp[ -i \tau  \sum_{i=1}^L g \sigma^x_i ]
\end{equation}
We take $\tau = 0.8$ and obtain an almost linear decrease for $\ln |A(t)|^2$ as shown in Fig.~\ref{fig:kim_huse}. This implies that $A(t)$ decays exponentially in time when the only conservation law (energy) is removed.


\section{Discussion}


We believe that the diffusive scaling of the R\'enyi entanglement entropy\cite{rakovszky_sub-ballistic_2019,huang_dynamics_2019} derives mainly from the unitary time-evolution operator of the system. This perspective leads us to propose the amplitude $A(t)$ in Eq.~\ref{eq:DefOfA}, which is the overlap between the unitary evolution operator of the entire system and that of the tensor product of the two decoupled right- and left- half systems. In a diffusive non-integrable system, we expect the amplitude $A(t)$ to decay as $e^{ - \sqrt{Dt} }$ (Eq.~\eqref{eq:A_proposal}), which directly implies an upper bound on the {\it operator} R\'enyi entropy $S_{n>1}^{\rm op}[U(t)]$ that scales as $\sqrt{t}$, as well as an upper bound on the annealed averaged R\'enyi entropy ${S^{(a)}_{n>1}(t)}$ for states.
For a spin-$\frac{1}{2}$ random local unitary circuit model with a $\text{U}(1)$ conservation law, we have been able to show analytically the bound $\overline{|A^2|} \ge e^{ - \sqrt{Dt} }$: In order to arrive at this result, we have mapped the dynamics of the disorder average to a diffusion process of $\zst$ and $\ost$ particles with a decimation process in the center of the system. This allowed us to show that the circuit average $\overline{|A|^2}$ can be bounded from below by the corresponding survival probability that scales as $e^{- \sqrt{Dt}}$. We have argued that the picture of $\ost$ and $\zst$  particles undergoing a classical diffusion process will also hold for a typical realization of the $U(1)$-symmetric circuit without random averaging, and likely also for generic diffusive non-integrable models with or without disorder. 

We can also discuss the amplitude $A(t)$ in a quantum field theory setting. One possible approach is to analytically continue this quantity to imaginary time, and compute
\begin{equation}
\tr(  e^{ -\sigma H } e^{ - \tau ( H_L + H_R ) } ) 
\end{equation}
where $\sigma = \frac{\beta}{2} + it$, $\tau = \frac{\beta}{2} - i t $. 
In a path integral formulation, one can think of this as a partition function on a cylinder
\footnote{generalized "cylinder" when the spatial dimension is larger than one.} 
with a slit of length $\tau$ at the center. For a conformal field theory in (1+1) dimensions, this represents the two point function of the twist field, which reads
\begin{equation}
\label{eq:A_beta_tau}
\begin{aligned}
  A( \beta, \tau  ) = \left( \frac{\pi}{\beta} \frac{1}{\sin \frac{\pi \tau }{\beta}}  \right)^{ 4h }, 
\end{aligned}
\end{equation}
where $h$ is the conformal weight (dimension) of the twist field. Going back to real time, we see that this amplitude decays as $A(t) \propto$ $e^{-4 h \frac{\pi }{\beta} t}$. With conformal symmetry, space and time are on an equal footing, which implies ballistic transport but excludes diffusive transport whose dynamical exponent is $z = 2$. Technically,  conformal symmetry also completely fixes the correlation function of the twist field evaluated in Eq.~\ref{eq:A_beta_tau} which, as shown above, leads to the
exponential decay of the amplitude $A(t)$. This is consistent with the fact that in a global quench scenario, a (1+1)-dimensional conformal field theory always has a linear growth of all the R\'enyi entropies. It would be interesting to understand the behavior of the amplitude $A(t)$, introduced in this paper, in other interacting quantum field theories, especially the ones that exhibit diffusive transport. 

{ For completeness, we briefly discuss $n$th R\'enyi entropies $S_n$ with $0 < n < 1$.
The R\'enyi entropies $S_{0<n < 1}(t)$ are generally not affected by the presence of a diffusive mode since they give higher weights to the smaller entanglement eigenvalues. The inequality $S_1(t) \le S_{0 < n < 1} (t) $ and the linear growth of the von Neumann entropy requires a not slower than linear growth of $S_{0<n< 1}(t)$. On the other hand, from the Lieb-Robinson bound\cite{lieb_finite_1972} for a Hamiltonian with local interactions, we expect that quantum information propagates ballistically with a velocity no faster than the Lieb-Robinson velocity. Hence $S_{0 < n < 1} (t)$ should still grow linearly.}

{ Finally, we compare our result for chaotic systems with possible diffusive entanglement scaling in integrable models. It is well known that the von Neumann as well as R\'enyi entanglement entropies will both
grow linearly when time-evolved under a  global quench by an integrable Hamiltonian 
(for a recent review, see e.g.  \cite{calabrese_quantum_2016}).
The linear growth behavior is ascribed to pairs of ballistically propagating quasi-particles, that could be interacting with each other\cite{alba_entanglement_2018}. We note that it
is possible to design diffusive dynamics for quasi-particles: To our
knowledge, one could at least design a CFT subject to a random collection of
conformal slits  in space-time that reflect the quasi-particle stochastically (the stochastic CFT)\cite{bernard_diffusion_2017} . Then all the R\'enyi entropies with R\'enyi index $n > 0$ will asymptotically grow as $\sqrt{t}$. The von Neumann entropy and 
the R\'enyi entropies $S_{0<n<1}$ make no exception because entanglement is ``transported'' by the quasi-particles and is thus completely determined by them. 
This is in sharp contrast to the case of chaotic models studied in this paper.}


\acknowledgements
TZ thanks Andrea De Luca, Benjamin Doyon, Tibor Rakovszky, Yichen Huang for discussions about the diffusive mode and proof techniques in general.
TZ was supported by postdoctoral fellowships from the Gordon and Betty Moore Foundation, under the EPiQS initiative, Grant GBMF4304, at the Kavli Institute for Theoretical Physics. 
This research is supported in part by the National Science Foundation under Grant No. NSF PHY-1748958.
We acknowledge support from the Center for Scientific Computing from the CNSI, MRL: an NSF MRSEC (DMR-1720256) and NSF CNS-1725797.
This work was supported in part by the NSF under Grant No. DMR-1309667 (AWWL).

\appendix
\section{Bound for the operator entanglement}
\label{app:op_EE_bd}

In this section, we prove the inequality in Eq.~\eqref{eq:A_le_max_lamb} stating that $|A(t)|$ is less than or equal to the square root of the maximal eigenvalue of the reduced {\it operator} density matrix of the time-evolution operator $U(t)$.

We will use the von Neumann inequality\cite{mirsky_trace_1975} which states the following: Let $X$ and $Y$ be complex $n\times n$ complex matrices, 
whose singular values \footnote{i.e. the square-roots of the eigenvalues of the non-negative Hermitian matrices $X X^\dagger$, and $Y Y^\dagger$, respectively.}, in non-increasing order, are denoted by
$\lambda^X_1 \geq \lambda^X_2 \geq ... \geq \lambda^X_n\geq 0$, and
$\lambda^Y_1 \geq \lambda^Y_2 \geq ... \geq \lambda^Y_n \geq 0$, respectively.

Then the von Neumann inequality is satisfied:
\begin{equation}
\label{eq:von-inequality}
|\tr( X Y ) | \le \sum_i \lambda^X_i\lambda^Y_i 
\end{equation}

We will take $X$ and $Y$ to be the matrices representing the operators $U(t)$ and $U_L(t)U_R(t)$,  respectively, in an orthonormal basis of operators, $\{V^{L}_i\}$ and$\{V^{R}_i\}$, of the left and the right subsystem.
We expand
\begin{align}
U(t)  &= \sum_{ij} X_{ij} V_i^L V_j^R  \\
U^{\dagger}_L(t)U^{\dagger}_R(t)  &= \sum_{ij} Y_{ij} V_i^L V_j^R  
\end{align}
Then according to the way we construct the operator reduced density matrix reviewed in Sec.~\ref{sec:implies_ee}, we have
\begin{equation}
(\lambda_i^X)^2 = \lambda^{\rm op}_i (t) 
\end{equation}

But since $U^{\dagger}_L(t)U^{\dagger}_R(t)$ is a single tensor product of operators on the left and right subsystems, the rank of $Y$ must be one and 
\begin{equation}
\lambda_1^Y = 1, \quad \lambda_{i>1}^Y = 0 
\end{equation}

Therefore according to Eq.~\eqref{eq:von-inequality}, we have
\begin{equation}
|A(t)| = |\tr( XY )| \le \lambda_1^X = \sqrt{\lambda^{\rm op}_{\rm max} (t)}
\end{equation}

The proof for Eq.~\eqref{eq:LambdaPsiMaxInequality} is similar. Now we take
\begin{equation}
\begin{aligned}
X &= U(t) |\psi(0) \rangle \langle \psi(0) | U^{\dagger} (t) \\ 
Y &= U_L(t)U_R(t) |\psi(0) \rangle \langle \psi(0) | U^{\dagger}_L(t)U^{\dagger}_R(t)
\end{aligned}
\end{equation}
and then obtain
\begin{equation}
\label{eq:trXYeq}
|\tr( XY)| = |\langle  \psi (0) | U_L^{\dagger} U_R^{\dagger}  U | \psi(0) \rangle|^2. 
\end{equation}

Since $|\psi(0)\rangle$ is a product state, the operator $Y$ defined above can be decomposed as a single tensor product of operators acting on the left and right subsystems. Hence
\begin{equation}
\lambda_1^Y = 1, \quad \lambda_{i>1}^Y = 0. 
\end{equation}
On the other hand, given the Schmidt decomposition of the state
\begin{equation}
U(t) |\psi(0)\rangle = \sum_\alpha \sqrt{\lambda^{\psi}_{\alpha}} | \alpha \rangle_L | \alpha\rangle_R 
\end{equation}
the Schmidt eigenvalues of the operator $X$ are the products $\sqrt{\lambda_\alpha^\psi}$ $\sqrt{\lambda_\beta^\psi}$.

Therefore, by Eq.~\eqref{eq:von-inequality}, we have
\begin{equation}
|\tr(XY)| \le \lambda_1^X = \text{max}_{\alpha,\beta}\sqrt{\lambda^{\psi}_{\alpha}\lambda^{\psi}_{\beta} } 
:= \lambda^{\psi}_{\rm max} 
\end{equation}
Combining with Eq.~\eqref{eq:trXYeq}, we obtain the inequality in Eq.~\eqref{eq:LambdaPsiMaxInequality}.


\section{Averages in the random circuit with a conserved charge}
\label{app:rand_aver_u1}

Moments of the matrix elements of the Haar random matrix can be computed analytically\cite{collins_integration_2006,gu_moments_2013}. In this paper we only need the second moment. For $d\times d$ unitary matrix $U$ drawn from the Haar ensemble, we have
\begin{equation}
\overline{U_{i_1j_1} U^*_{i_2 j_2}} = \frac{1}{d}\delta_{i_1 i_2} \delta_{j_1 j_2} 
\end{equation}

In the text, we use a block diagonal ${\rm U}(1)$ conserved matrix as the building block of the circuit, where each block is an independent Haar random matrix. Consequently, the second moment is only nonzero when the average is performed within the same block. Because of different block sizes, we have 
\begin{equation}
\label{eq:UU_aver}
\overline{U_{i_1j_1} U^*_{i_2 j_2}} = \delta_{i_1 i_2} \delta_{j_1 j_2} 
\left\lbrace
\begin{aligned}
  & 1 & \quad  i_1 = j_1 = \up \up \text{ or }  \dn \dn \\
  & \frac{1}{2} & \quad  i_1, j_1 \in \{\up \dn, \dn \up \}  \\
\end{aligned} \right. 
\end{equation}

The tensor $\delta_{i_1 i_2}$ in the result indicates that the gate will only produce $| \zst \rangle $ and $|\ost \rangle $ states. We then translate Eq.~\eqref{eq:UU_aver} to the symmetric simple exclusion process in Eq.~\eqref{eq:ssep}. 

For the unpaired center gate, we have
\begin{equation}
\overline{U_{i_1j_1}} = 0
\end{equation}
which produces the decimation in Eq.~\eqref{eq:decimation}.


\bibliographystyle{apsrev4-1}
\bibliography{diffusion_EE_paper}

\end{document}